\documentclass[a4paper,11pt]{article}
\pdfoutput=1 

\usepackage{jcappub} 

\usepackage[T1]{fontenc} 
\usepackage[dvipsnames]{xcolor}
\usepackage{tikz}

\usepackage{amssymb, amsmath, epsfig, natbib}

\author[a,b]{Martin White}

\affiliation[a]{Department of Physics, University of California,
Berkeley, CA 94720}
\affiliation[b]{Department of Astronomy, University of California,
Berkeley, CA 94720}

\emailAdd{mwhite@berkeley.edu}

\title{A marked correlation function for constraining
       modified gravity models}


\abstract{Future large scale structure surveys will provide increasingly tight
constraints on our cosmological model.  These surveys will report results on
the distance scale and growth rate of perturbations through measurements of
Baryon Acoustic Oscillations and Redshift-Space Distortions.  It is interesting
to ask: what further analyses should become routine, so as to test
as-yet-unknown models of cosmic acceleration?  Models which aim to explain
the accelerated expansion rate of the Universe by modifications to General
Relativity often invoke screening mechanisms which can imprint a non-standard
density dependence on their predictions.  This suggests density-dependent
clustering as a `generic' constraint.  This paper argues that a density-marked
correlation function provides a density-dependent statistic which is easy to
compute and report and requires minimal additional infrastructure beyond what
is routinely available to such survey analyses.  We give one realization of
this idea and study it using low order perturbation theory.  We encourage
groups developing modified gravity theories to see whether such statistics
provide discriminatory power for their models.}
\arxivnumber{1609.08632}

\begin{document}
\maketitle
\flushbottom

\section{Introduction}

The observation that the expansion rate of the Universe is accelerating
is one of the most puzzling aspects of our current cosmological model.
Two classes of explanation have been investigated, one based on a modification
of the contents of the Universe \cite{Wei13,PDG14}
and one based on a modification of gravity
(see e.g.~Refs.~\cite{JaiKho10,Cli12,Joy15,Hut15} for recent reviews).
At present there are no theoretically consistent, observationally allowed
models which provide cosmic acceleration through modifications to gravity.
Therefore, observational constraints on modifications to general relativity
(GR) often focus on `generic' features that some as-yet-to-be-determined
models might be expected to have.

Large-scale structure surveys typically provide constraints on the
distance scale and rate-of-growth of fluctuations through studies of
baryon acoustic oscillations (BAO) and redshift-space distortions (RSD)
\cite{Wei13,PDG14}.
In combination with gravitational lensing surveys (of galaxies or the
cosmic microwave background) a number of tests of GR on linear scales can
be constructed \cite{Wei13,PDG14}.
It is reasonable to assume that such analyses will be an integral part of
analyses of future surveys as well, improving some tests of dark energy and
modified gravity models.
A question then arises what other analyses should `routinely' be performed
on future surveys, so that observational constraints are available for
theorists and phenomenologists seeking to constrain next-generation models?
Ideally these analyses should be simple to perform and report, while at
the same time providing information beyond the standard analyses
currently published.

In the absence of a specific theoretical framework this is a difficult
question to answer.  The tightest constraints will come from a model-by-model
analysis, but more generic constraints can also be useful when investigating
wide classes of models.  One frequently encountered phenomenon in modified
gravity models is a screening mechanism that forces model predictions to
approach those of GR in regions of high density or strong gravitational
potential.  Conversely, signatures of modified gravity will show up in regions
where gravity is weak.  Screening is a property of many modified gravity
models that offers potentially distinctive observational signatures.

Here we advocate the use of the density-marked correlation function
\cite{WhiPad09}
as an easy-to-compute statistic which may test future modified gravity models.
Computation of the marked correlation function requires minimal modification
to existing analysis frameworks, and requires no further infrastructure beyond
that which is routinely available for studying BAO and RSD.
By weighting the pairs of galaxies by a `mark' which depends on a local
density estimate (e.g.~increasing the weight of low density regions) it
provides density-dependent information from the survey which may be of use
in constraining future theories.

This paper introduces the inverse-density-marked correlation function and
provides an exploration of its properties using low order Lagrangian
perturbation theory.  The latter is primarily for convenience -- such
statistics can also be calculated and explored using N-body simulations
or mock catalogs which will give access to smaller scales where the signal
may be larger.  We shall return to this point in the conclusions.
The outline of this paper is as follows.
In section \ref{sec:background} we introduce the marked correlation
function and quickly review Lagrangian perturbation theory and how to
compute the marked correlation function within this framework.
We present some results to build intuition on the marked correlation
function in section \ref{sec:results}.
Finally we conclude in section \ref{sec:conclusions}.

\section{Background}
\label{sec:background}

This section presents some background to set the stage and our notation.
Our focus will be the prediction of the density-marked two-point function
in redshift space, i.e.~the marked correlation function.
For technical reasons, we shall restrict ourselves here to the angle
averaged (or monopole moment of the) correlation function though further
information would almost certainly be contained in the higher moments.

\subsection{Marked correlation function}

The marked correlation function
\cite{BeiKer00,BeiKerMec02,Got02,SheTor04,SCS05,Ski06}
is a generalization of the usual, configuration space, 2-point function.
If each object is assigned a mark, $m$, the marked correlation function
is defined as
\begin{equation}
  \mathcal{M}(r) \equiv \frac{1}{n(r)\bar{m}^2}\sum_{ij} m_i m_j
  = \frac{1+W}{1+\xi}
\end{equation}
where the sum is over all pairs of a given separation, $r$, $n(r)$ is the
number of such pairs and $\bar{m}$ is the mean mark for the entire sample.
The second equality serves to define $W$ and emphasizes that we expect
$\mathcal{M}\to 1$ at large scales.
In general the marked correlation function is very easy to evaluate if
one is already able to evaluate the `normal' correlation function.  Its
computation can be trivially integrated into a standard clustering pipeline
with almost no overhead, and it can make use of the same masks, mock catalogs
and codes as the standard analysis.
Aside from a measure of density\footnote{At least one estimate of the density
is often computed as part of the analysis pipeline in order to do BAO
reconstruction \cite{Eis07}.}, it requires no additional survey products
beyond those traditionally used for configuration-space clustering analyses.

The use of local density as a mark was suggested in Ref.~\cite{WhiPad09}.
In what follows we shall be interested in marks which are functions
of a smoothed density field, $\rho_R$.
There are numerous ways of estimating a density given a collection
of objects
(see e.g.~\cite{Sch00,Deh01,Eis03,AscBin05,Rom07,Pad12,FalKoyZha15}
 for some examples).
The scale over which the density field can be estimated is tied to the
mean separation of objects, which for current and future surveys aimed
at BAO or RSD is likely to be close to linear (see Appendix \ref{app:scale}).
If we assume\footnote{Obviously there is no need to make this approximation
in the data or if the prediction is generated from mock catalogs.  It is made
purely to facilitate a perturbative treatment of $\mathcal{M}(r)$.} that the
density which is used as a mark is in fact the linear density field,
smoothed on some scale $R$, then it is straightforward to make an
analytic prediction for the marked correlation function within Lagrangian
perturbation theory.

The choice of mark is in principle arbitrary.  The smaller the range of the
mark the smaller the signal, but the more stable the result.  Marks with a
very large range can introduce noise.  In general smoothly varying marks are
preferred over very rapidly changing ones.
If we had a particular signal we were looking for it would be possible to
optimize the mark -- for example we could use a value based on the screening
maps introduced in Ref.~\cite{Cab12} or on probability of being in a sheet
or filament (e.g.~Refs.~\cite{BonKofPog96,Col07,Alp15,Alo15,FalKoyZha15};
Refs.~\cite{Fal14,FalKoyZha15} argue that some screening mechanisms are
sensitive to the dimensionality of the surrounding structure).
Highly complex marks could be simulated, but are unlikely to be analytically
tractable.  Marks which can be expressed in terms of the local Lagrangian
density and its low order derivatives can be easily handled within Lagrangian
perturbation theory (e.g.~using the techniques described in
Ref.~\cite{VlaCasWhi16}).

In the absence of a specific theoretical target, we choose simple functions
of a smoothed density field.
As an example we shall consider marks of the form
\begin{equation}
  m = \left(\frac{\rho_\star+1}{\rho_\star+\rho_R}\right)^p
\end{equation}
where $\rho_R$ is our smoothed density in units of the mean density,
$\bar{\rho}$, and $\rho_\star$ is a parameter we can adjust
(an alternative would be $\exp[-\rho_R/\rho_\star]$).
The contribution of low density regions can be enhanced by choosing $p>0$.
In general one would compute several of these marked correlation functions,
obvious examples would be fixing $p$ and varying $\rho_\star$ over some
range or fixing $\rho_\star$ and varying $p$.
Note that upweighting the low density regions is similar in spirit to
computing the properties of voids, often suggested as a probe of modified
gravity (e.g.~Refs.~\cite{Cla13,Ham14,Ham15,Cai15,Ziv15})
but without the need to find voids and characterize their purity and
completeness.
It is another way to study clustering as a function of environment and a
generalization of a void-galaxy cross correlation
(e.g.~Refs.~\cite{Cai16,Ham16,AchBla16,Fag16}
for recent studies in observations and simulations respectively).
Like the void probability function, the marked correlation function involves
an infinite tower of higher order correlation functions, though only a small
fraction of the possible configurations.
The marked correlation function is similar to statistics of the `clipped'
density field \cite{Sim11}, which Ref.~\cite{Lom15} advocated as a test of
modified gravity, or to the log-transformed density field \cite{NeySzaSza09}.
As such, the analytic development below may also be applicable to some limits
of clipped or log-transformed correlation functions.

It will prove useful below to Taylor expand $m$ in powers of the smoothed
density contrast, $\delta_R$,
\begin{equation}
  m \simeq 1 - \frac{p}{1+\rho_\star}\,\delta_R 
             + \frac{p(p+1)}{2(1+\rho_\star)^2}\,\delta_R^2 + \cdots
\label{eqn:Bn}
\end{equation}
If $\rho_\star\gg 1$ the mark is slowly varying with $\delta$ and the
expansion converges quickly.  As we lower $\rho_\star$, or increase $p$,
we require more terms to adequately describe the mark.

\subsection{Lagrangian perturbation theory}

The Lagrangian approach to cosmological structure formation was developed
in \cite{Zel70,Buc89,Mou91,Hiv95,TayHam96,Mat08a,Mat08b,CLPT,Mat15} and
traces the trajectory of an individual fluid element through space and time.
It has been well developed in the literature, and extended to include
effective field theory corrections and a generalized bias model
\cite{PorSenZal14,VlaWhiAvi15,VlaCasWhi16}.
In this exploratory foray we will use the first order solution, linear in
the density field (also known as the Zeldovich approximation \cite{Zel70})
and a simple local Lagrangian bias scheme.
It is plausible, though by no means proven, that a low order expansion may
work better for a statistic that emphasizes low density regions than for a
statistic focused on high density peaks.  In any event, the extension of the
theory to higher orders, and to more complex biasing schemes, is
straightforward and can be attempted if it proves useful.

We shall closely follow the notation in Refs.~\cite{Mat08a,WanReiWhi14}, in
particular we define
\begin{equation}
\langle\delta(\mathbf{q}_1)\delta(\mathbf{q}_2)\rangle=
 \langle\delta_1\delta_2\rangle_c = \xi(\mathbf{q}=\mathbf{q}_1-\mathbf{q}_2)
 \quad ,
\end{equation}
\begin{equation}
\Delta_i=\Psi_i(\mathbf{q}_1)-\Psi_i(\mathbf{q}_2) \qquad , \qquad
A_{ij} = \langle\Delta_i\Delta_j\rangle_c \qquad {\rm and} \qquad
U_i = \langle\delta_1\Delta_i\rangle_c
\end{equation}
with $\Psi_i$ the Lagrangian displacement (i.e.~the final position $\mathbf{x}$
is related to the initial position $\mathbf{q}$ of a fluid element or dark
matter particle by $\mathbf{x}=\mathbf{q}+\mathbf{\Psi}(\mathbf{q})$).
The two-point function of the density is the usual correlation function,
$\xi$, which is a function only of the (Lagrangian) separation $\mathbf{q}$.
The other two 2-point functions are the auto-correlation of the relative
Lagrangian displacement, $\Delta_i$, which we denote $A_{ij}$ and the
cross-correlation of the displacement and density, $U_i$.  These are also
only functions of the separation, $\mathbf{q}$.

We shall assume that our tracers have local Lagrangian bias \cite{Mat08a}
\begin{equation}
  1+\delta(\mathbf{x},t) = \int d^3q\ F\left[\delta_L(\mathbf{q})\right]
  \delta_D\left[\mathbf{x}-\mathbf{q}-\mathbf{\Psi}(\mathbf{q},t)\right]
\end{equation}
specified by $F$.  We shall describe $F$ via its low-order bias expansion,
with $b_1=\langle F'\rangle$ and $b_2=\langle F''\rangle$ the Lagrangian
bias parameters \cite{Mat08a,Mat08b}.

As discussed above we assume that our mark can be expressed as a function
of the smoothed, linear theory density field (an extension to derivatives
of this smooth field, e.g.~shear, is straightforward in principle).
We shall denote the smoothed overdensity as $\delta_R$ and
write
\begin{equation}
U^R_i=\langle\delta_{R,1}\Delta_i\rangle_c \qquad , \qquad
\xi_{R,1}=\langle\delta_1\delta_{R,2}\rangle_c \qquad {\rm and}\qquad
\xi_R=\langle\delta_{R,1}\delta_{R,2}\rangle_c \quad .
\end{equation}
The zero-lag correlators can be written as $\sigma^2_{R,1}$ and $\sigma^2_R$.
Writing the mark as $G[\delta_R]$ and performing the same bias expansion as
for $F$ with coefficients $B_1$ and $B_2$ we have the mean mark
$\bar{m}=1+b_1B_1\sigma_{R,1}^2+\cdots$.

The marked correlation function can now be computed using standard methods
\cite{Mat08a,Mat08b,CLPT,Whi14}.  Let us define $\Xi$ such that
\begin{equation}
  \mathcal{M}(r) = \frac{1+W}{1+\xi}
                 = \frac{\Xi(b_i,B_i)}{\Xi(b_i,B_i\equiv 0)}
\end{equation}
Up to third order in the bias expansion (and second order in clustering)
\begin{eqnarray}
    \Xi &=&\int d^3q\ \dfrac{1}{(2\pi)^{3/2}|A|^{1/2}}
    e^{-(1/2)(q_i-r_i)(A^{-1})_{ij}(q_j-r_j)}
    \nonumber \\
    &\times& \bigg\{ 1 + b_1^2 \xi_L  - 2 b_1 U_ig_i
    - [b_2 + b_1^2] U_iU_jG_{ij}
    - 2 b_1 b_2 \xi_L U_ig_i
    + \cdots
    \nonumber \\
    && + B_1^2 \xi_R  - 2 B_1 U_i^Rg_i
    - [B_2 + B_1^2] U_i^RU_j^RG_{ij}
    - 2 B_1 B_2 \xi_R U^R_ig_i
    + \cdots
    \nonumber \\
    && + 2b_1B_1 \left(\xi_{R,1}  - 2 U_iU_j^RG_{ij}\right)
       -2B_1^2b_1g_iU_i\xi_R - 2b_1^2B_1g_iU^R_i\xi 
    \nonumber \\
    && -2(b_2+b_1^2)B_1g_iU_i\xi_{R,1}
       -2(B_2+B_1^2)b_1g_iU^R_i\xi_{R,1}
       + \cdots \bigg\}
\label{eqn:xi-kernel-m}
\end{eqnarray}
where $g_i=(A^{-1})_{ij}(q_j-r_j)$ and $G_{ij}=(A^{-1})_{ij}-g_ig_j$
(not to be confused with our mark function, $G[\delta_L]$).

\begin{figure}
\begin{center}
\resizebox{\columnwidth}{!}{\includegraphics{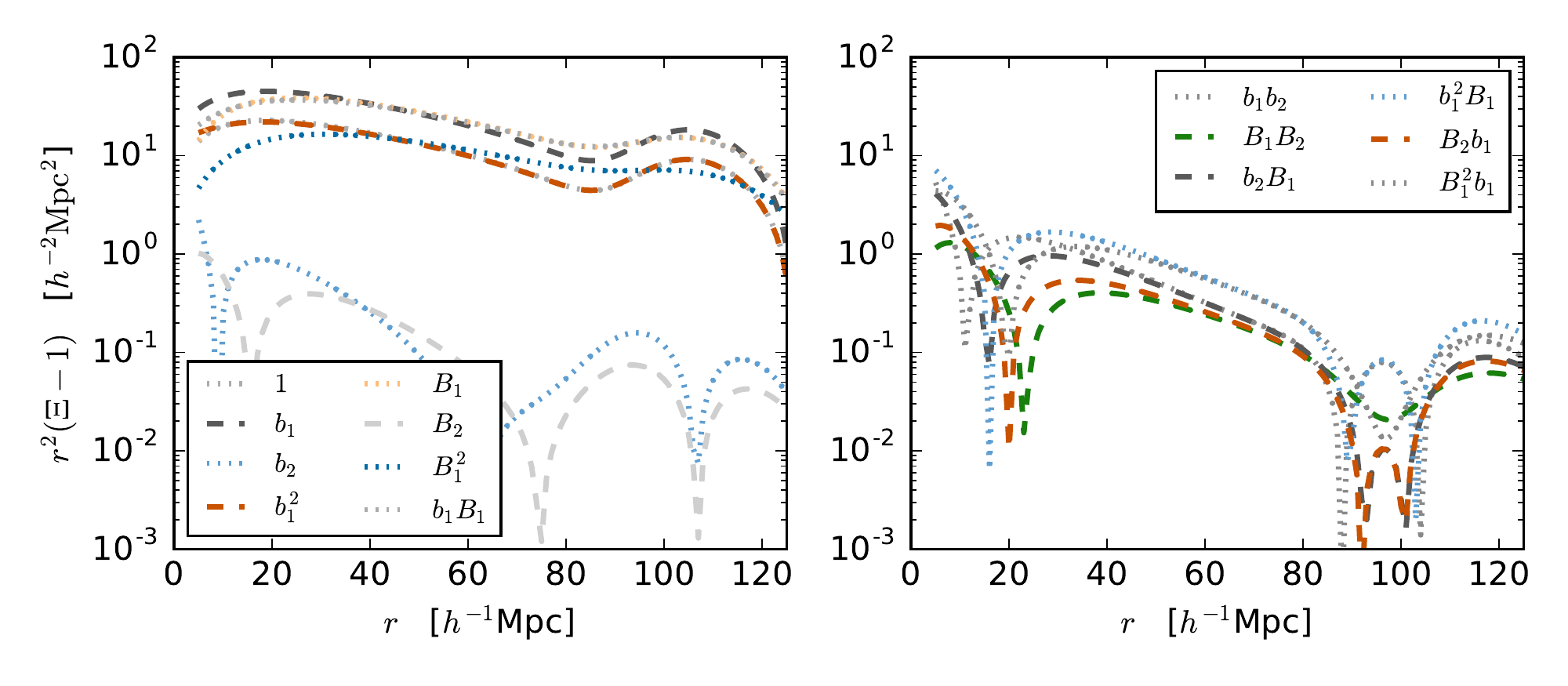}}
\end{center}
\caption{The contributions to $\Xi$, in real space, as a function of $r$.
We have assumed a Gaussian smoothing with a width of $10\,h^{-1}$Mpc to
define $\rho_R$ and a $\Lambda$CDM cosmology at $z=0.55$ to define the
linear theory power spectrum.
(Left) The first 8 contributions, up second order in the bias expansion.
The $1$ and $b_1^2$ terms are nearly indistinguishable.
(Right) The $3^{\rm rd}$ order terms, which are generally small on large
scales.}
\label{fig:xir}
\end{figure}

Fig.~\ref{fig:xir} shows the contributions to $\Xi$ for a $\Lambda$CDM
cosmology at $z=0.55$.
We have defined the density, $\rho_R$, appearing in our mark using a
Gaussian of width $10\,h^{-1}$Mpc.
Fig.~\ref{fig:xir} demonstrates that the sum is dominated on large scales
by terms involving $1$, $b_1$ and $B_1$, with the $b_2$ and $B_2$ terms
being significantly smaller.
Keeping only the contribution from the $1$ term in the $\{\cdots\}$ gives
the Zeldovich approximation for the matter correlation function.
On large scales this looks like a smoothed version of the linear theory
correlation function \cite{Bha96}.
Including non-zero $b_1$ describes biased tracers, with the standard,
large-scale, Eulerian bias $b=1+b_1$.  The terms involving $B_i$ are the
terms which characterize the density dependence of the mark (i.e.~the
$1^{\rm st}$ and $2^{\rm nd}$ derivatives of $G[\delta_L]$ with respect
to $\delta_L$) so $B_1<0$ describes a mark which emphasizes lower densities.
The $b_1^2$ and $B_1^2$ terms also look very much like a smoothed version of
the linear theory correlation function on large scales (arising as they
do from an almost-convolution of $\xi$ with the Gaussian in
Eq.~\ref{eqn:xi-kernel-m}).  The $b_1$ and $B_1$ terms have a similar shape
to the ``1'', but an amplitude about twice as large (for $b_1=B_1=1$)
\cite{Mat08b,CLPT,Whi14}.
These facts will help to explain the behavior we will see below in the
marked correlation function.

In redshift space we define the angle-averaged $\mathcal{M}$ by first
averaging the numerator and denominator over angle, and then performing
the division.
This is different from averaging the ratio, but serves to make the denominator
the monopole of the redshift space correlation function.
With this definition, to go into redshift space we simply multiply the
line-of-sight components of $U$ and $A$ by $1+f$ before doing the $d^3q$
integral, then perform the division \cite{Mat08b,Whi14}.
As the Zeldovich approximation does a relatively good job of describing the
monopole of the redshift-space correlation function but a poor job on the
higher moments \cite{Whi14} we shall restrict attention to this angle-averaged
statistic here.  Then the relative sizes of the terms do not change much
between real- and redshift-space and Fig.~\ref{fig:xir} remains a good guide
to the structure of the theory.

\section{Results}
\label{sec:results}

\begin{figure}
\begin{center}
\resizebox{\columnwidth}{!}{\includegraphics{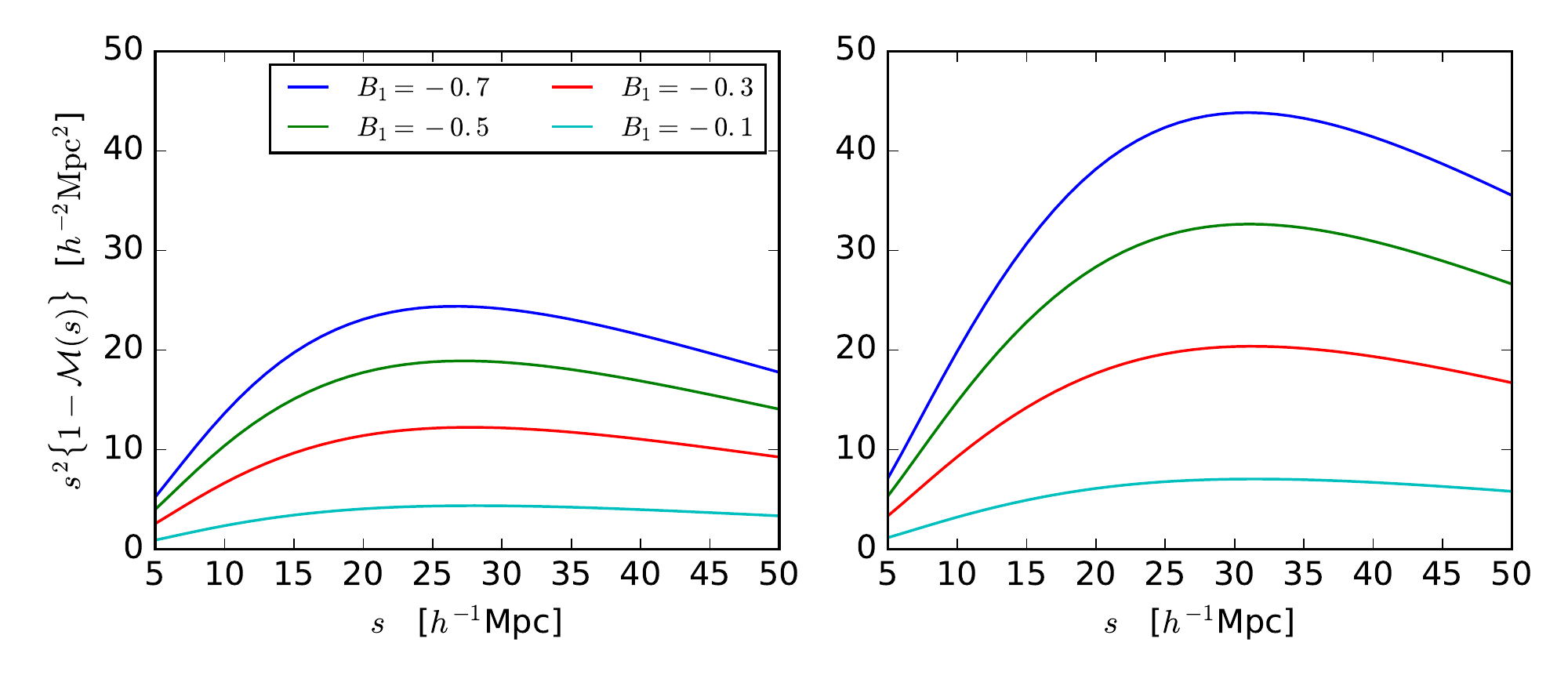}}
\end{center}
\caption{The monopole of the marked correlation function for unbiased
($b_1=b_2=0$; left) and biased ($b_1=1$, $b_2=0$; right) tracers assuming
$B_2=B_1^2$.
We plot $s^2(1-\mathcal{M})$ to allow the use of a linear $y$-axis scale
and because on large scales we expect $1-\mathcal{M}\propto\xi_0$ (see text).}
\label{fig:mark}
\end{figure}

With the formalism in hand we can now explore the behavior of $\mathcal{M}$
as we change the bias of the tracer and the mark.
To begin we consider the marked correlation function for an unbiased tracer
of the density field ($b_1=b_2=0$) with the same definition of density as
above.  Motivated by Eq.~(\ref{eqn:Bn}) we shall take $B_1<0$ and $B_2=B_1^2$.
Fig.~\ref{fig:mark} (left) shows the angle-averaged marked correlation
function at $z\simeq 0.55$ for three values of $B_1$.  Given the structure
in Fig.~\ref{fig:xir} and recalling that $B_1<0$ we see $M<1$.
To enhance the dynamic range we have plotted $s^2[1-\mathcal{M}]$,
which we see is qualitatively similar to $r^2\xi$ as might be expected by
the structure of the terms (see below).

To gain some intuition let us consider just the lowest order term, linear
in $B_1$ (still with $b_1=b_2=0$).  This term is $-2B_1U^R_ig_i$.
Recalling that $U^R_i=\langle\delta_R\Delta_i\rangle$ we see that this term
describes the density dependence of the (infall) velocity in GR
(and Lagrangian PT).
On large scales the $-U_ig_i$ contribution is very similar to $\xi_L$
\cite{CLPT,Whi14}, so this term makes $\mathcal{M}-1$ look like
$(1+2B_1)\xi_L$.
If the density dependence of infall velocity is modified from its GR form
we would expect to see a departure in the measured dependence of $\mathcal{M}$
on $B_1$ (through changes in $\rho_\star$ or $p$) on the scales where this
modification manifests.
Knowing the $B_1$ and $r$-dependence of this modification would provide
valuable information as to the form of the departure from GR.
If the linear correlation function, $\xi_L$, is not affected by the
departure from GR, then we expect the $U$ effect to dominate on large
scales.  Only on smaller scales the other terms, e.g.~$G_{ij}U_iU_j$
describing more complex density-velocity-velocity correlations, become
comparable in size to the $U_i$ terms (Fig.~\ref{fig:xir}).

As a second example we consider biased tracers.  For definiteness we have
taken $b_1=1$ (corresponding to a large-scale, Eulerian bias of 2) and
$b_2=0$ (though the results are quite insensitive to $b_2$).
Fig.~\ref{fig:mark} shows that the structure of $\mathcal{M}$ is quite
similar to the unbiased case, but the amplitude is increased.
A future survey such as DESI\footnote{{\tt http://desi.lbl.gov}} or
Euclid\footnote{{\tt http://sci.esa.int/euclid}},
covering many Gpc${}^3$ of volume, would be able to measure $\mathcal{M}$ with
per cent level precision on scales of many Mpc.

Our analytic calculation explains the major trends we see here.
For large $s$, $\mathcal{M}\simeq 1+W-\xi$.
On scales above the smoothing scale, $R$, the contributions
$\xi\simeq\xi_{R,1}\simeq\xi_R$ and $U^R\simeq U$.
Further, the contributions from $-U_ig_i$ are almost the same as those from
$\xi$, as mentioned above.
Thus on large scales $\Xi\simeq 1+(1+b_1+B_1)^2\xi_0$ and
\begin{equation}
  1-\mathcal{M}(s) \simeq \xi-W \simeq -B_1(2+2b_1+B_1)\xi_0 + \cdots
\end{equation}
This gives a simple prediction for the scaling of the amplitude with the
mark (e.g.~choice of $\rho_\star$) and the bias of the tracer, valid on
large scales and to low order in perturbation theory.

If the theory of gravity is modified, in a density dependent way, we would
expect the ingredients going into this calculation to also be modified.  We
mentioned above the example of $U_i=\langle\delta\Delta_i\rangle$, describing
the correlation between velocities and densities.
This could be modified from its form in GR if there are additional,
density-dependent forces at work.
Precise prediction of the functional form of any deviations would require a
calculation within a specific model, and the scale at which these departures
occur is likely to be model dependent.
With a suitably flexible bias model, low order perturbation theory (such as
presented above) suffices for GR predictions on large scales.
To probe for departures occuring on smaller scales corrections to this model
must be computed.
Higher order terms in the bias expansion and higher order terms
in the dynamics can be systematically computed if desired.
This would allow us to push to quasi-linear scales, but not to fully
non-linear scales.  A model for non-linear clustering would need to be
derived from numerical simulations.
This is straightforward, given a suitable model for the halo occupancy of
the galaxies.  Mock catalogs can also include the effects of masking or
missing data, complex marks and other survey non-idealities.

\section{Conclusions}
\label{sec:conclusions}

The growth of large-scale structure as observed in modern cosmological
surveys offers one means of testing general relativity on the largest
scales.  Constraints on the distance scale, growth rate and deflection of
light (from BAO, RSD and lens modeling respectively) have become standard
and such analyses are likely to be performed on all future surveys.
In the absence of a single, compelling model of modified gravity it is
difficult to know how to augment these `standard' analyses so as to best
constrain modifications.

Many models which modify gravity invoke a screening mechanism that forces
the predictions to become equal to those of GR in regions of high density
or strong gravitational potential.  This suggests that generic tests for
the density dependence of the growth of structure could be added to our
list of `standard' analyses and might provide useful in constraining models
of modified gravity in the future.
In this paper we have pointed out that an inverse-density-marked correlation
function, $\mathcal{M}(r)$, is very easy to compute and report, requiring
little computational overhead, code development or additional survey products.

To illustrate the form that $\mathcal{M}(r)$ takes in the standard theory,
we present the lowest order Lagrangian perturbation theory calculation.
In keeping with our perturbative approach we have emphasized large scales,
such as can be easily probed with upcoming surveys aimed at BAO and RSD.
Of course it may be that modifications to gravity are more easily seen on
smaller scales with density fields estimated from denser samples of galaxies
(or other objects).  While this likely invalidates the perturbative
calculation presented earlier, it is relatively straightforward to generate
predictions on such scales from simulations given a suitably refined model
for bias (e.g.~populating halos in N-body simulations with galaxies using an
HOD model \cite{CooShe02} or a semi-analytic model \cite{Bau06}).
Indeed, even if such statistics prove unconstraining for the modified gravity
models of the future, the density information they encode can be very valuable
for validating and refining the bias model used to model BAO and RSD
statistics, e.g.~breaking degeneracies \cite{WhiPad09} or testing for assembly
bias.
Should the population of galaxies in a survey depend on properties of a
dark matter halo beyond its mass (e.g.~formation time or concentration) which
correlates with large-scale environment we expect to see a departure from the
simplest theoretical models which neglect such effects.  The nature of this
departure can be understood from the theory above and parameterized in a very
flexible way.  Since the GR predictions are so well known, as long as
modifications to gravity are not fully degenerate with these effects,
we should still be able to disentangle them.

There are several obvious lines of development.  First, this statistic could
be computed on existing modified gravity simulations to get a sense for the
size of the effect as a function of scale and the ideal redshift and number
density of tracer for this test.  This would help to calibrate expectations,
but cannot be taken as definitive since existing models are either ruled out
or do not explain acceleration with modified gravity and we do not know how
the predictions would differ in a model which could explain our Universe.
Second, a study should be undertaken of the best density estimate and to
what extent noise in this estimate adversely affects the results.
Third, if further investigation warrants, this model can be extended to
higher order in perturbation theory or to include a dependence on the
derivatives of the density or the dimensionality of the structure.
Within the Lagrangian framework it is relatively straightforward to include
marks which can be expressed in terms of initial density and its derivatives.
Finally, it is worth investigating whether marked statistics for auto- and
cross-correlations of imaging and spectroscopic surveys could yield other,
valuable constraints on modifications to GR.  It is straightforward
\cite{Mat08b,CLPT,WanReiWhi14} to modify the formulae in this paper to
account for cross-correlations of biased and marked tracers, thus opening
the possibility to predict a range of other statistics.

\acknowledgments
I would like to thank
Nikhil Padmanabhan for early conversations about this project and
Joanne Cohn, Shirley Ho, Bhuvnesh Jain, Uros Seljak and Miguel Zumalacarregui
for useful comments on a draft of this manuscript.

This work was begun at the Aspen Center for Physics, which is supported
by National Science Foundation grant PHY-1066293.
I thank the Center for its hospitality. 
This research has made use of NASA's Astrophysics Data System.
The analysis in this paper made use of the computing resources of the
National Energy Research Scientific Computing Center.

\appendix

\section{The smoothing scale}
\label{app:scale}

In the text we have assumed that the smoothed density field, estimated from
the galaxies, can be reasonably well approximated by the linear density field.
In this appendix we provide a back-of-the-envelope argument that this is
likely to be true for surveys which are designed to study BAO or RSD.

Most surveys focused on large-scale structure adjust their sampling so that
$\bar{n}P$ is of order a few, where $P$ is the power spectrum evaluated at a
convenient scale, often $0.2\,h\,{\rm Mpc}^{-1}$.
If we assume that the smoothing scale that defines the density in our mark,
$\rho$, scales with the mean interobject separation, $\bar{n}^{-1/3}$,
this implies $\rho$ is close to linear.

As an example, imagine $P(k)=P_\star\,(k_\star/k)$ for some fiducial $k_\star$.
This is approximately the slope of the CDM power spectrum on quasi-linear
scales today.  Further imagine we smooth the field with a filter of size
$R=r\bar{n}^{-1/3}$ with $r\sim 1$.  We have
\begin{equation}
  \sigma^2(R) = \int\frac{k^2\,dk}{2\pi^2} P(k)W^2(k;R)
              = \frac{k_\star^3\,P_\star}{2\pi^2} \int d\kappa
                \ \kappa\,W^2(\kappa;k_\star R)
\end{equation}
with $\kappa=k/k_\star$.  For a Gaussian of width $R$
\begin{equation}
  \sigma^2(R) = \frac{k_\star^3\,P_\star}{2\pi^2}\ \frac{1}{2(k_\star R)^2}
\end{equation}
To simplify this, let us choose $k_\star$ so that $k_\star^3 P_\star/2\pi^2=1$.
This means $k_\star$ is where the dimensionless power is unity.
Now if we say $\bar{n}P_\star=\nu$ then
\begin{equation}
  k_\star R=\left(\frac{2\pi^2}{P_\star}\right)^{1/3}\ r\bar{n}^{-1/3}
           \simeq \frac{2.7\,r}{\nu^{1/3}}
\end{equation}
and hence
\begin{equation}
  \sigma(R) = \frac{\nu^{1/3}}{2^{5/6}\pi^{2/3}\,r} = 0.26\,\frac{\nu^{1/3}}{r}
\end{equation}
This suggests that for modest $\nu$ and $r\ge 1$ a kernel density estimate
should return an approximately linear density field.
If the galaxy field is highly biased, then the matter field which it represents
will be even closer to linear.

\bibliographystyle{JHEP}
\bibliography{ms}

\end{document}